\documentclass[letterpaper, 12pt]{article}
\usepackage{graphicx}
\usepackage{float}
\usepackage{hyperref}
\usepackage{lineno}

\newcommand\pubdate{\today}

\def\Title#1{\begin{center} {\Large #1 } \end{center}}
\def\Author#1{\begin{center}{ \sc #1} \end{center}}
\def\Address#1{\begin{center}{ \it #1} \end{center}}

\newcommand\pubblock{\rightline{\begin{tabular}{l}  \\ 
         \pubdate  \end{tabular}}}
\newenvironment{Abstract}{\begin{quotation}  }{\end{quotation}}
\newenvironment{Presented}{\begin{quotation} \begin{center} 
             PRESENTED AT\end{center}
      \begin{center}\begin{large}}{\end{large}\end{center} \end{quotation}}

\begin{document}
    \begin{titlepage}
        \pubblock
        \Title{Transverse Single-Spin Asymmetries of Midrapidity $\pi^{0}$ and $\eta$ Mesons in $\sqrt{s_{NN}} = 200$ GeV $p^{\uparrow}+$Au and $p^{\uparrow}+$Al Collisions from PHENIX }
        \Author{Dillon Fitzgerald for the PHENIX Collaboration}
        \Address{University of Michigan, Ann Arbor}
        \vfill
        \begin{Abstract}
            Understanding the spin structure of the proton is of large interest to the nuclear physics community and it is one of the main goals of the spin physics program at the Relativistic Heavy Ion Collider (RHIC). Measurements from data taken by the PHENIX detector with transverse ($p^{\uparrow} + p$, $p^{\uparrow}$ + Al, $p^{\uparrow}$ + Au) proton polarization play an important role in this, in particular, due to the leading order access to gluons in polarized protons. Transverse single-spin asymmetries (TSSAs) provide insight into initial and final state spin-momentum and spin-spin parton-hadron correlations. In addition to possible final state contributions, $\pi^{0}$ and $\eta$ TSSAs access both quark and gluon correlations in the polarized proton. Furthermore, the $p^{\uparrow} + A$ data from RHIC provides an opportunity to study the effect of TSSAs in the presence of additional nuclear matter. Midrapidity $\pi^{0}$ and $\eta$ mesons are measured at PHENIX by detecting the 2$\gamma$ decay with the electromagnetic calorimeter (EMCal) in the central arm spectrometer, which has fine granularity for the resolution of separate decay photons. New results for TSSAs of midrapidity $\pi^{0}$ and $\eta$ mesons in $\sqrt{s_{NN}} = 200$ GeV $p^{\uparrow}$ + Au and $p^{\uparrow}$ + Al collisions from the 2015 running year will be presented, and compared with the recent $\sqrt{s} = 200$ GeV $p^{\uparrow} + p$ results. \end{Abstract}
    
        \vfill
        \begin{Presented}
            DIS2023: XXX International Workshop on Deep-Inelastic Scattering and
            Related Subjects, \\
            Michigan State University, USA, 27-31 March 2023 \\
            \includegraphics[width=9cm]{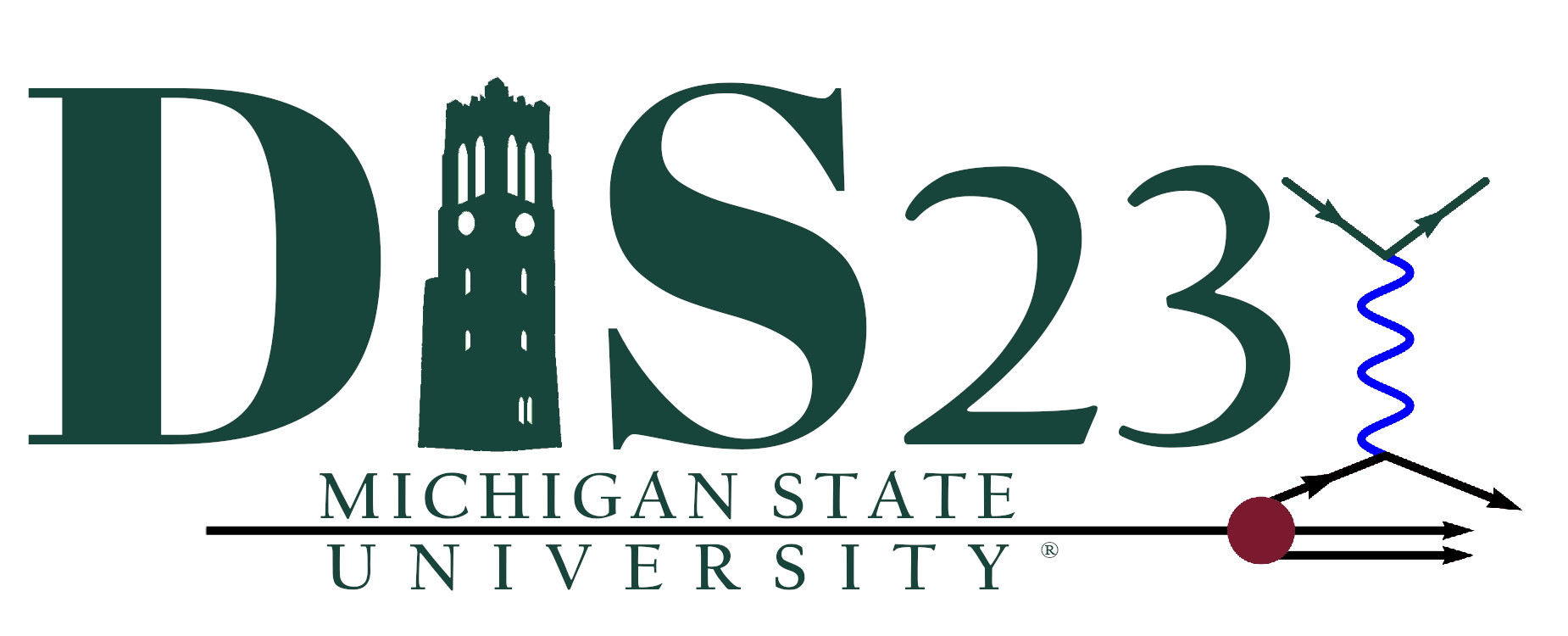}
        \end{Presented}
    \end{titlepage}

    \section{Introduction}
        The complex nature of hadronic bound states lead to spin-momentum and spin-spin correlations analogous to the fine and hyperfine structure in atomic bound states. Transverse single-spin asymmetries (TSSAs) provide insight into these correlations~\cite{TSSA_review}, with measurements at hadron-hadron colliders providing leading-order access to gluons. TSSAs measure azimuthal modulations of produced particles in collisions involving polarized hadrons, and are defined as
        \begin{equation}
            A_{N} (\phi) = \frac{\sigma^{\uparrow}(\phi) - \sigma^{\downarrow} (\phi)}{\sigma^{\uparrow}(\phi) + \sigma^{\downarrow} (\phi)} = A_{N} \cos\phi,
            \label{eq:AN_phi}
        \end{equation}
        \noindent where $\sigma^{\uparrow, \downarrow}(\phi)$ correspond to transversely polarized cross sections for different spin orientations. Negligible contributions to such observables were predicted from perturbatively calculable hard scattering coefficients ( $<1\%$)~\cite{pQCDTSSA}. While more recent calculations suggest possible modest contributions at 2 loops~\cite{twoLoopsTSSA}, this is still far less than the large magnitudes of observed TSSAs across several collision energies and experiments~\cite{Klem:1976ui, Adams:1991cs, Allgower:2002qi, Arsene:2008aa}. 

        \begin{figure}[H]
            \includegraphics[width=0.95\textwidth]{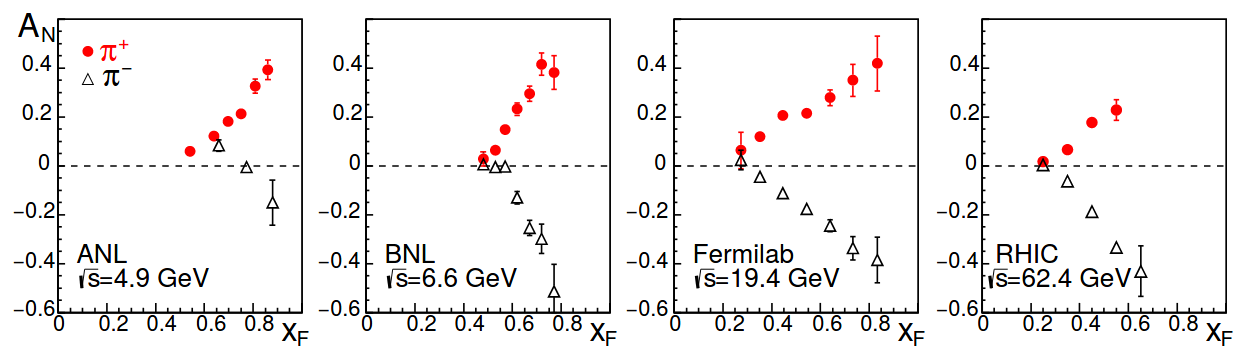}
            \caption{Transverse single-spin asymmetries for $\pi^{+}$ (red circles) and $\pi^{-}$ (black triangles) as a function of $x_{F} = 2p_{z}/\sqrt{s}$ for several different experiments with different collision energies. The asymmetries grow as large as $40\%$ in magnitude as $x_{F}$ increases, with opposing sign for $\pi^{+}$ and $\pi^{-}$.}
            \label{fig:pipm_ANxF}
        \end{figure}

        This implies that the mechanism behind the observed TSSAs is nonperturbative. Two theoretical frameworks have been developed in which the nonperturbative elements of the asymmetry calculations lead to the large TSSAs observed in nature: (1) Higher twist effects and (2) Transverse momentum dependent functions. In the former, power-suppressed terms by the hard scale $Q$ in the factorization expansion are considered, so the collinear factorization scheme is preserved, and only a hard scale is needed for the measurement for which the $p_{T}$ of the measured particle is taken as a proxy. In the latter, standard collinear parton distribution functions (PDFs) and fragmentation functions (FFs) are extended to depend on the transverse momentum of partons within hadrons or in the process of hadronization. In the TMD approach, both a hard scale ($Q$) and a soft scale ($k_{T}$) must be measured with sufficient scale separation ($Q>>k_{T}$). Recent TSSAs measurements in polarized $p+p$ collisions at the Relativistic Heavy Ion Collider (RHIC) and the possible mechanisms contributing to them can be found in Refs.~\cite{PPG246,PPG247,PPG235,PPG234,PPG238,STAR:2022hqg,STAR:2020nnl,STAR:2017wsi,STAR:2017akg}. The TSSAs here are presented as a function of $p_{T}$, making the higher twist formalism applicable in this case.

        In $p^{\uparrow}+A$ collisions, the underlying partonic origins of the asymmetry could be affected by the presence of a heavy nucleus, making it particularly interesting to compare TSSAs in various collision systems and rapidity regions. In a factorized picture, the heavy nucleus should have no effect on initial-state spin-momentum correlations leading to observed asymmetries, while final-state spin-momentum correlations can be affected due to the transport through more nuclear matter and a more complex QCD environment. On the other hand, in a picture where factorization is broken, interactions with the nuclear remnant may have some effect on the observed asymmetry~\cite{fact_break_2010,extra_asymmetries}. Recently the theoretical community has proposed comparing forward hadron TSSAs in $p^{\uparrow}+p$ and $p^{\uparrow}+A$ collisions to investigate the interplay between spin physics and low-x physics, in particular, to probe gluon saturation effects in the nucleus~\cite{Kang:2011ni}. These type of measurements have been performed at RHIC for forward charged hadrons~\cite{PPG215, PPG226}, and forward $J/\psi$ mesons~\cite{PPG211} by the PHENIX experiment and forward neutral pions by the STAR experiment~\cite{STAR:2020grs}. Strong nuclear dependence was observed in the case of forward charged hadron production by PHENIX, while only a moderate dependence was observed by STAR for the $\pi^{0}$ measurement, despite it being even farther forward in rapidity. Far forward neutron TSSAs were also measured by PHENIX~\cite{PPG203,PPG244}, where nuclear dependence was observed and understood to be due to the interplay of electromagnetic interactions in ultra-peripheral collisions and hadronic interactions. These proceedings, following the work in Ref.~\cite{PPG204}, show the first measurement of midrapidity TSSAs in $p^{\uparrow}+A$ collisions for both $\pi^{0}$ and $\eta$ production. The measurements are also compared with those for midrapidity $\pi^{0}$ and $\eta$ production from $p^{\uparrow}+p$ collisions from Ref.~\cite{PPG234}, revealing no observed nuclear modification in $p^{\uparrow}+A$ collisions. 

    \section{Analysis Methods}
\begin{figure}[H]
            \includegraphics[width=0.99\textwidth]{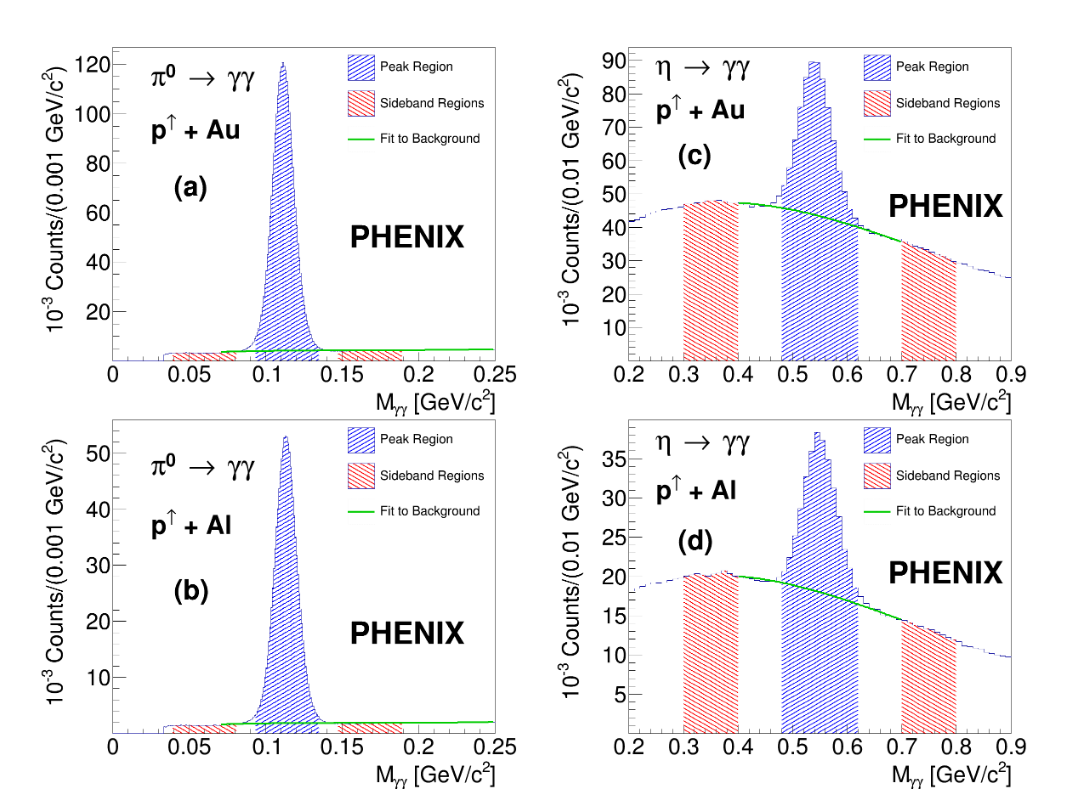}
            \caption{Invariant mass distributions around the $\pi^{0} \rightarrow \gamma\gamma$ peak in (a) $p^{\uparrow}$ + Au collisions and (b) $p^{\uparrow}$ + Al collisions and around the $\eta \rightarrow \gamma\gamma$ peak in (c) $p^{\uparrow}$ + Au collisions and (d) $p^{\uparrow}$ + Al collisions for photon pairs within $4<p_T~[{\rm GeV}/c]<5$ in the west central-arm spectrometer from Ref.~\cite{PPG204}. The blue leftward-hatched regions are the signal peaks, used for quantifying yields for the $A^{N}$ calculations, the red rightward-hatched regions are the side bands, used to quantify yields for the $A_{N}^{\rm BG}$ calculations, and the green solid curves correspond to fits to the combinatorial background, used in calculating the background fractions ($r$).}
            \label{fig:invmass}
        \end{figure}

        The data for the measurement were taken in 2015 at RHIC with total integrated luminosities of approximately 202 and 690 nb$^{-1}$ for $\sqrt{s_{NN}} = 200$ GeV $p^{\uparrow}$ + Au and $p^{\uparrow}$ + Al collisions, respectively. Transverse single-spin asymmetries were measured for both $\pi^{0} \rightarrow \gamma \gamma$ and $\eta \rightarrow \gamma \gamma$ production at midrapidity ($| \eta | < 0.35$) using the electromagnetic calorimeters (EMCal) in the central arm spectrometers at PHENIX. An EMCal trigger was used in coincidence with a minimum bias trigger to select high $p_{T}$ photons for this analysis. In addition, tracks measured by the drift and pad chambers in the central arm spectrometer that are associated with an EMCal cluster are vetoed to remove background from charged particles. For more information regarding the PHENIX detector subsystems relevant for this measurement, and the selection criteria placed on the $\pi^{0} \rightarrow \gamma \gamma$ and $\eta \rightarrow \gamma \gamma$ candidates, see Ref.~\cite{PPG204}, where the measurement is described in detail.  Figure ~\ref{fig:invmass} shows the diphoton invariant mass distributions around the $\pi^{0}$ and $\eta$ peaks for photon pairs within $4<p_T<5$ GeV/$c$ in the west central arm spectrometer in $p^{\uparrow}$ + Au and $p^{\uparrow}$ + Al collisions.  
        
        Once the $\pi^{0} \rightarrow \gamma \gamma$ and $\eta \rightarrow \gamma \gamma$ samples are curated, TSSAs can be calculated from yields shown in the blue leftward hatched region of Fig.~\ref{fig:invmass} using the ``relative luminosity'' formula, which is equivalent to Eq.~\ref{eq:AN_phi} with additional correction factors for the proton beam polarization fraction and the azimuthal detector coverage of the central arm spectrometers at PHENIX. It is defined as
        \begin{equation}
            A_N = \frac{1}{P\left<|\cos\phi|\right>} \frac{{N^{\uparrow}}-\mathcal{R}{N^{\downarrow}} }{{N^{\uparrow}}+\mathcal{R}{N^{\downarrow}}},
            \label{eq:relan}
        \end{equation}
        where $P$ is the beam polarization, and $\mathcal{R}$ is the relative luminosity, defined as the ratio of integrated luminosities between the bunches with $\uparrow$ and $\downarrow$ spin states. $\left<|\cos\phi|\right>$ is the azimuthal acceptance factor  of the detector, calculated separately for each $p_{T}$ bin and spectrometer arm, and $N$ refers to the yields, with the arrows referring to the up ($\uparrow$) or down ($\downarrow$) polarization of the proton beam. The measured asymmetries were corrected for background as follows, 
        \begin{equation}
            A_N^{\rm sig} = \frac{A_N-r \cdot A_N^{\rm BG}}{1-r},
            \label{eq:sb}
        \end{equation}
        where $A_{N}$ is the asymmetry calculated in the signal region with Equation~\ref{eq:relan}, $r$ is the background fraction under the $\pi^{0}$ or $\eta$ peaks, calculated from the background fits, and $A_N^{\rm BG}$ is the background asymmetry, which was evaluated in the side bands on both sides of the $\pi^{0}$ and $\eta$ peaks. The background asymmetry was consistent with zero across collision systems and particle species for the entire measured $p_{T}$ range. Details on cross-checks and systematic studies can be found in Ref.~\cite{PPG204}.   

    \section{Results}
\begin{figure}[H]
            \includegraphics[width=0.99\textwidth]{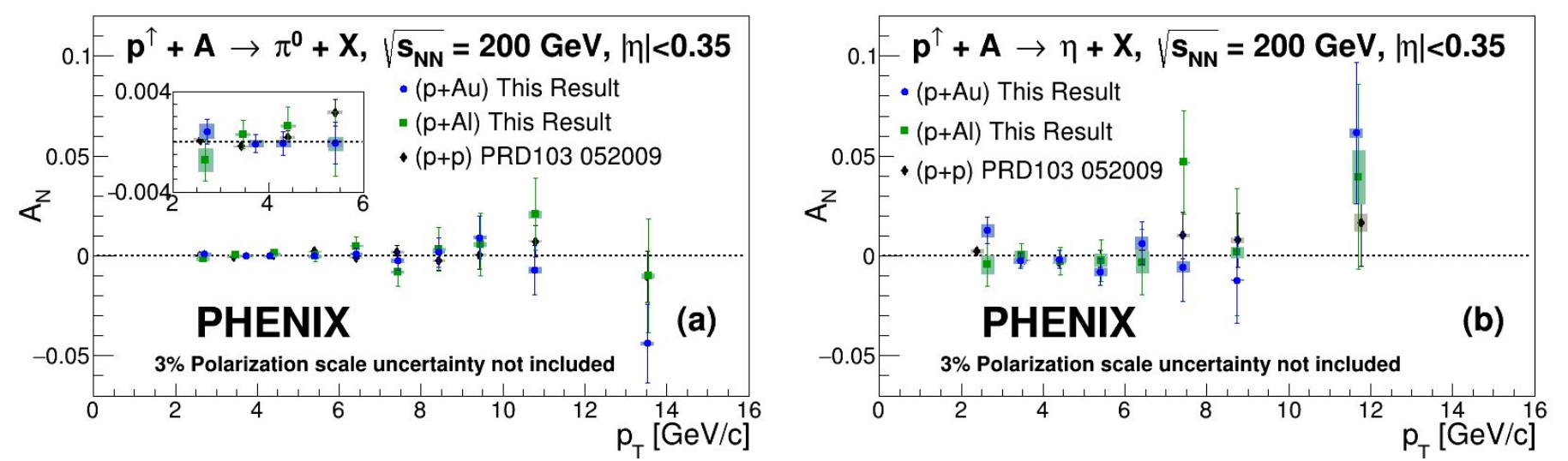}
            \caption{Transverse single-spin asymmetry for (a) $\pi^{0}$ and (b) $\eta$ mesons in $p^{\uparrow}$ + Au collisions (blue circles), and $p^{\uparrow}$ + Al collisions (green squares) from Ref.~\cite{PPG204}, shown alongside the same measurement in $p^{\uparrow} + p$ collisions from Ref.~\cite{PPG234} (black diamonds). The error bars represent the statistical uncertainty ($\sigma^{\rm stat}$) while the boxes represent the total systematic uncertainty ($\sigma^{\rm syst}$).}
            \label{fig:pi0eta_AN}
        \end{figure}

        The TSSA ($A_{N}$) for midrapidity $\pi^{0}$ and $\eta$ mesons are shown in Fig.~\ref{fig:pi0eta_AN}, with the $\pi^{0}$ results shown in panel (a), and the $\eta$ results shown in panel (b). The asymmetries for $p^{\uparrow}$ + Au and $p^{\uparrow}$ + Al collisions from Ref.~\cite{PPG204} are plotted with blue circles and green squares respectively. The $p^{\uparrow}+A$ results are plotted alongside the $p^{\uparrow} + p$ results from Ref.~\cite{PPG234}, shown in black diamonds for comparison. All asymmetries in Fig.~\ref{fig:pi0eta_AN} are consistent with zero, meaning there is no nuclear modification of the TSSA observed at midrapidity. 

    \section{Summary}
        TSSA measurements in $p^{\uparrow}+A$ collisions using the 2015 RHIC data have yielded surprises and led to many open questions in the spin physics community. Presented were the results from the first midrapidity TSSA measurement from the 2015 RHIC $p^{\uparrow}+A$ data, in particular, for midrapidity ($| \eta | < 0.35$) $\pi^{0} \rightarrow \gamma \gamma$ and $\eta \rightarrow \gamma \gamma$ production in $\sqrt{s_{NN}} = 200$ GeV $p^{\uparrow}$+Au and $p^{\uparrow}$+Al collisions~\cite{PPG204}. This measurement serves as an important milestone in our understanding of the modification of such observables in the presence of additional nuclear matter. 


\end{document}